\documentclass[preprint,prb,superscriptaddress]{revtex4}

\usepackage{graphicx}
\usepackage{dcolumn}
\usepackage{bm}
\usepackage{color}

\begin{document}
\title{Critical Oxide Thickness for Efficient Single-walled Carbon Nanotube Growth on Silicon Using Thin SiO$_2$ Diffusion Barriers}

\author{J. M. Simmons}
 \affiliation{Department of Physics, University of Wisconsin - Madison}
\author{Beth M. Nichols}
 \affiliation{Department of Chemistry, University of Wisconsin - Madison}
\author{Matthew S. Marcus}
 \affiliation{Department of Physics, University of Wisconsin - Madison}
 \affiliation{Department of Chemistry, University of Wisconsin - Madison}
\author{O. M. Castellini}
 \affiliation{Department of Physics, University of Wisconsin - Madison}
\author{R. J. Hamers}
 \affiliation{Department of Chemistry, University of Wisconsin - Madison}
\author{M. A. Eriksson}
 \affiliation{Department of Physics, University of Wisconsin - Madison}

\begin{abstract}
The ability to integrate carbon nanotubes, especially single-walled
carbon nanotubes, seamlessly onto silicon would expand the range of
applications considerably.  Though direct integration using chemical
vapor deposition is the simplest method, the growth of single-walled
carbon nanotubes on bare silicon and on ultra-thin oxides is greatly
inhibited due to the formation of a non-catalytic silicide.  Using
x-ray photoelectron spectroscopy, we show that silicide formation
occurs on ultra-thin oxides due to thermally activated metal
diffusion through the oxide.  Silicides affect the growth of
single-walled nanotubes more than multi-walled nanotubes due to the
increased kinetics at the higher single-walled nanotube growth
temperature.  We demonstrate that nickel and iron catalysts, when
deposited on clean silicon or ultra-thin silicon dioxide layers,
begin to form silicides at relatively low temperatures, and that by
900~$^{\text{o}}$C, all of the catalyst has been incorporated into
the silicide, rendering it inactive for subsequent single-walled
nanotube growth.  We further show that a 4~nm silicon dioxide layer
is the minimum diffusion barrier thickness which allows for
efficient single-walled nanotube growth.
\\

\\
*Corresponding author:\\
M. A. Eriksson\\
Department of Physics,
University of Wisconsin - Madison\\
1150 University Ave.\\
Madison, WI  53706\\
e-mail:  maeriksson@wisc.edu
\end{abstract}

\maketitle

\section{\label{sec:level1}Introduction}
Since their discovery, carbon nanotubes have shown great promise for
a wide variety of applications which utilize their unique electronic
and mechanical properties.$^[$\cite{Saitobook}$^]$
 \normalsize For applications in which individual nanotubes act as
the working element, such as nanotube field effect transistors
(FETs)$^[$\cite{MartelAPL,TansNature}$^]$ or chemical
sensors,$^[$\cite{ChopraAPL,KongScience,LeeNanoLett}$^]$ it is
important to control the location and orientation of the nanotubes.
Nanotubes can be prefabricated and then assembled into the desired
geometry, or they can be fabricated in place using chemical vapor
deposition (CVD).$^[$\cite{DaiJPCB,KongCPL}$^]$  CVD is preferred
since the growth can be widely tuned, both in yield and structure
(single- vs. multi-walled), by modifying the experimental
conditions.  While there are a vast number of CVD recipes available
in the literature for both single- and multi-walled nanotube growth,
most studies of the growth process have focused on the role of the
carbon precursor (e.g. CO, CH$_4$, C$_2$H$_2$) and temperature as
control parameters with less attention placed on the choice of
catalyst and substrate.$^[$\cite{MoisalaJPCondMatt}$^]$ Optimization
of the catalytic process requires an understanding of the catalyst
chemistry throughout the growth process, including the initial
chemical state of the catalyst before the introduction of the
feedstock, as well as the catalytic decomposition of the feedstock
in or on the catalyst
particle.$^[$\cite{BakerJCatal,dlArcosJPCB,dlArcosAPL,KlinkeJPCB,KonyaPCCP,EmmeneggerCarbon,PrabhakaranLang}$^]$
Only recently have studies focused on the catalyst-feedstock and
catalyst-substrate
interaction.$^[$\cite{KonyaPCCP,EmmeneggerCarbon,PrabhakaranLang,dlArcosCPL,HerreraJCatal,NishimuraJJAP,YangAPL}$^]$

It is well established that the yield of single-walled nanotube
growth on silicon substrates is dramatically reduced compared with
growth on thick silicon dioxide layers, due to poisoning of the
catalyst by the formation of a
silicide.$^[$\cite{dlArcosAPL,JungNanoLett,MaedaPhysicaE,SohnAPL}$^]$
Intriguingly, the growth of multi-walled nanotubes on silicon
substrates is regularly reported and seems less susceptible to
catalyst
poisoning,$^[$\cite{CaoAPL,CheungPNAS,KondoJJAP,WongAPL}$^]$ yet the
cause of this difference has not been addressed. For device
applications it is desirable to use thin oxides to increase the gate
capacitance and gate efficiency, and thus it is important to
understand how to minimize oxide thickness while still preventing
catalyst poisoning. Though thick silicon dioxide layers have been
used as diffusion barriers during nanotube growth, there has been no
investigations to date that determine the the effectiveness of
ultra-thin oxides.

Here we demonstrate explicitly that catalyst diffusion through the
ultra-thin silicon dioxide layer controls the formation of the
non-catalytic silicide. Using x-ray photoelectron spectroscopy
(XPS), we study the interfacial reactions between the substrate and
iron- or nickel-based nanotube catalysts during the initial
temperature ramp portion of a CVD growth cycle.  We show that
ultra-thin oxide layers (4~nm or greater) are sufficient to inhibit
the silicide formation and permit high yield growth of single-walled
carbon nanotubes.  On thinner oxides or clean silicon, silicide
formation begins by 600~$^{\text{o}}$C (Ni) or 800~$^{\text{o}}$C
(Fe) and the catalyst is entirely consumed in the silicide at the
growth temperature of single-walled nanotubes. The silicide
formation temperatures account for the difference in single- versus
multi-walled nanotube growth because multi-walled nanotubes are
grown at lower temperatures where some of the catalyst remains
unreacted and active for catalysis. Interestingly, the silicides
form while the silicon dioxide is still present on the substrate,
indicating that diffusion of metal or silicon through the oxide is
occurring.  We show that it is metal diffusion through the oxide,
forming a metal silicide underneath the oxide, which dominates the
interfacial reactions between the nanotube catalyst and the silicon
substrate.

\section{\label{sec:level1}Results and Discussion}
\begin{figure}[htp]
 \centering
 \includegraphics[width=2.25in]{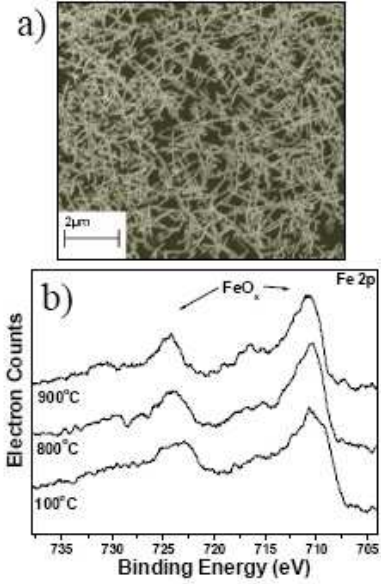}
 \caption{(a) SEM image of high-yield nanotube growth on 100~nm SiO$_2$.
 (b) Fe~2p core level spectra of iron nitrate catalyst on 100~nm SiO$_2$.
 As the sample is annealed, there is no appreciable change in the oxidation state of the iron.}
 \label{Fig1}
\end{figure}

To understand the catalyst-substrate interfacial reactions, we first
analyze XPS spectra of iron nitrate catalyst that has been deposited
on a thick (100~nm) thermal oxide. On thick oxides, silicides will
not form and single-walled nanotube growth occurs with high yield
(Figure~\ref{Fig1}a).
Before annealing, the iron core level (Figure~\ref{Fig1}b) exhibits
broad, asymmetric peaks at 710.8~eV and 724.5~eV, corresponding to
the Fe~2p$_{3/2}$ and 2p$_{1/2}$ levels of iron oxide in a mixed
Fe$^{2+}$/Fe$^{3+}$ oxidation
state.$^[$\cite{McIntyreAnalChem,LinApplSurfSci}$^]$ The extra
intensity on the high binding energy side of the core levels is due
to unresolved satellites that are characteristic of oxidized
iron.$^[$\cite{LinApplSurfSci,GrosvenorSurfInterfaceAnal}$^]$  As
the catalyst is heated on the thick oxide, there are no major
changes in the chemistry of the iron oxide. There is a small loss in
iron intensity, due to either agglomeration of the catalyst on the
surface (thereby reducing the measured intensity due to the finite
electron escape depth) or desorption of the iron into the vacuum
while annealing.$^[$\cite{KlinkeJPCB}$^]$ There is also a shift in
the center of gravity of the Fe~2p$_{3/2}$ core level as the
Fe$^{2+}$ oxide (FeO) transforms into the more stable Fe$^{3+}$
oxide (Fe$_2$O$_3$)$^[$\cite{NIST-JANAF}$^]$ and the satellite peaks
at 715~eV and 730~eV become better defined. Even at
1000~$^{\text{o}}$C, the catalyst remains as iron oxide because
there is no reaction pathway to reduce it to pure iron.  The
necessary reduction to metallic iron in a CVD reactor would come
from flowing hydrogen or from feedstock decomposition products,
enabling the catalysis of carbon nanotube growth.

At the opposite extreme - that of hydrogen-terminated silicon on
which single-walled nanotube growth is inhibited - surface reactions
play a major role in the evolution of catalyst chemistry even at low
temperatures.
\begin{figure}[htp]
 \centering
 \includegraphics[width=2.25in]{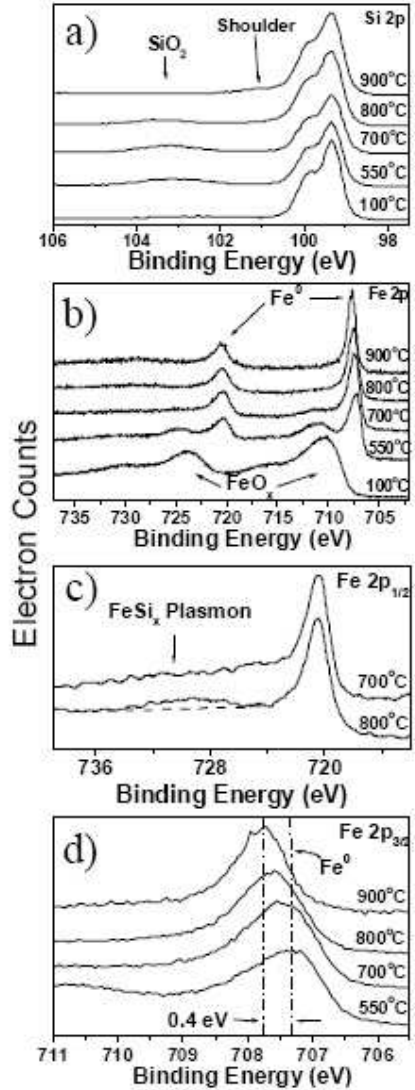}
 \caption{XPS spectra of iron nitrate on clean silicon, for the (a) Si~2p and (b) Fe~2p core levels.
 At low annealing temperatures, the iron is reduced by the silicon substrate, forming a thin layer of SiO$_2$.
 By 800~$^{\text{o}}$C, the iron begins to alloy with the substrate, forming an iron silicide, seen in
 (d) by a 0.4~eV shift in the Fe~2p$_{3/2}$ core level and in (c) by the emergence of a plasmon loss peak.}
 \label{Fig2}
\end{figure}
While iron nitrate catalyst that has been deposited on a hydrogen
terminated silicon substrate remains oxidized at room temperature
(Figure~\ref{Fig2}b), the center of gravity of the Fe~2p$_{3/2}$
core level is shifted to 710.5~eV, indicating that the iron has been
partially reduced as compared to the catalyst on the thick oxide.
Whereas the Si~2p spectrum of the thick oxide shows only a peak at
103.3~eV due to the SiO$_2$ layer, the spectrum of the
hydrogen-terminated substrate (Figure~\ref{Fig2}a) at room
temperature resolves the Si~2p atomic core levels (99.3 and
103.3~eV), as well as a small, broad peak on the high binding energy
side of the Si~2p peak, at around 102.7~eV, indicating the formation
of a non-stoichiometric silicon
suboxide.$^[$\cite{OrlowskiJAlloysCmpd}$^]$ This suboxide could have
formed in air after the HF oxide strip but before introduction in
the vacuum chamber, or could account for the small amount of
reduction in the iron as compared with the catalyst on the thick
oxide. Reports have indicated that a silicon dioxide is more stable
than either FeO or Fe$_2$O$_3$, which could lead to an exchange of
oxygen from the iron oxide to the silicon
substrate,$^[$\cite{NIST-JANAF,HommaJPCB}$^]$ even at low
temperature.

In contrast to the thick oxide, iron catalyst deposited on hydrogen
terminated silicon begins to reduce to metallic iron when the
catalyst by the time it is annealed to 550~$^{\text{o}}$C.  This is
seen in figure~\ref{Fig2}b by the decrease in Fe~2p intensity at
710.5~eV, 724.1~eV, and the associated satellites of iron oxide, and
by the appearance of peaks at 707.4~eV and 720.7~eV corresponding to
the Fe~2p core levels of Fe$^0$.  The iron oxide is further reduced
as the substrate is annealed, indicated by the reduced intensity of
the iron oxide peaks (Figure~\ref{Fig2}b).  By 800~$^{\text{o}}$C,
the iron oxide has been fully reduced. As in the case of a thick
oxide, the reduction in UHV must come from a surface reaction
between the oxidized catalyst and the substrate. Indeed, there is an
appreciable exchange of oxygen from the iron oxide to the silicon
substrate, leading to a marked increase in intensity of the silicon
oxide peak, accompanied by a shift in the peak position to higher
binding energies (Figure~\ref{Fig2}a).  At 550~$^{\text{o}}$C the
silicon oxide peak shifts to higher binding energy, and between
700~$^{\text{o}}$C and 800~$^{\text{o}}$C the peak reaches 103.3~eV
which is characteristic of stoichiometric SiO$_2$. This indicates
that the iron oxide catalyst is reduced by the silicon substrate
leading to the formation of a thin SiO$_2$ layer.

After the iron has been fully reduced at 800~$^{\text{o}}$C, the
catalyst begins to form a silicide.  At 800~$^{\text{o}}$C, the
Fe~2p$_{3/2}$ core level has shifted to higher binding energy by
approximately 0.2~eV (Figure~\ref{Fig2}d), and a very broad peak
begins to appear in the Fe~2p spectrum centered around 729~eV
(Figure~\ref{Fig2}c).  In the silicon core levels, a small shoulder
appears at 100.5~eV on the high binding energy side of the Si~2p
peak (Figure~\ref{Fig2}a).  By 900~$^{\text{o}}$C, the Fe~2p$_{3/2}$
peak has shifted further to 707.8~eV, giving a total binding energy
increase of 0.4~eV (Figure~\ref{Fig2}d). The 0.4~eV total shift in
the Fe~2p$_{3/2}$ peak position, accompanied by the appearance of a
peak $\sim$21~eV higher in binding energy, is indicative of the
formation of an iron silicide, likely
FeSi$_2$.$^[$\cite{RuhrnschopfTSF}$^]$ The peak at 729~eV is due to
a plasmon loss, seen in Figure 2c, and has been observed in electron
energy loss spectroscopy$^[$\cite{ZhuJAP}$^]$ as well as
XPS$^[$\cite{dlArcosAPL,RuhrnschopfTSF}$^]$ investigations of Fe/Si
multilayers. Further, the intensity of the shoulder on the Si~2p
peak increases and the SiO$_2$ has desorbed entirely as evidenced by
the loss of the $\sim$103.3~eV feature in the Si~2p spectrum. Though
the shoulder in the Si~2p spectrum at 100.5~eV seems to be
associated with the formation of the silicide, the apparent binding
energy of the shoulder is too high to be due to a FeSi$_x$ core
level which would be expected to be shifted by only
$\sim$0.2~eV.$^[$\cite{RuhrnschopfTSF}$^]$ At present, the origin of
this shoulder is not known. Annealing at 1000~$^{\text{o}}$C (not
shown) only leads to a decrease in the intensity of the iron peaks,
but causes no noticeable changes in the peak positions, indicating
that the silicide formation is complete at 900~$^{\text{o}}$C.

Catalyst poisoning by silicide formation is not unique to iron.  New
also see that nickel thin film catalyst on clean silicon is
incorporated into a silicide, only at a much lower temperature.
\begin{figure}[htp]
 \centering
 \includegraphics[width=2.25in]{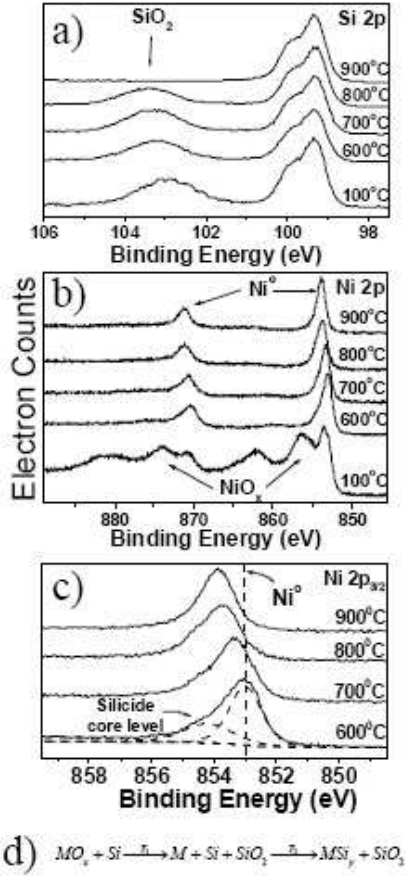}
 \caption{XPS spectra of a nickel film on clean silicon for the (a) Si~2p and (b) Ni~2p core levels.
 Similar to the iron catalyst, the nickel is first reduced by through the formation of more silicon dioxide,
 and then forms a silicide before reaching the single-wall nanotube growth temperature.
 (c) The nickel silicide begins to form at 600~$^{\text{o}}$C, seen as an up-shift in the Ni~2p core level.
 By 800~$^{\text{o}}$C, the core level has shifted by 1.0~eV, indicating that all of the nickel has been incorporated into a silicide.
 (d) Schematic reaction pathway from metal oxide catalyst to the formation of the non-catalytic silicide.}
 \label{Fig3}
\end{figure}
Whereas iron catalyst is not fully reduced until 800~$^{\text{o}}$C,
the native nickel oxide has been fully reduced by
600~$^{\text{o}}$C, and is accompanied by the growth of a SiO$_2$
layer (Figure~\ref{Fig3}a,b).  By 600~$^{\text{o}}$C some of the
nickel has also been incorporated into a silicide, seen as a high
binding energy shoulder on the Ni~2p$_{3/2}$ core level
(Figure~\ref{Fig3}c). As the substrate is further annealed, the
Ni~2p$_{3/2}$ level shifts to higher binding energy, reaching a
total shift of 1.0~eV by 800~$^{\text{o}}$C where the entire nickel
film has been incorporated into the
silicide.$^[$\cite{CheungJVST,NguyenPhysScripta}$^]$ The chemical
progression for both iron and nickel catalyst can be summarized as
in Figure 3d. The initial metal or metal oxide is completely reduced
by the silicon substrate leading to the growth of a silicon dioxide
layer. After being reduced, the metal can then react with the
silicon substrate directly, leading to the formation of a silicide.

The initial silicide formation at $\sim$600~$^{\text{o}}$C for
nickel and $\sim$800~$^{\text{o}}$C for iron explains the differing
growth yields of single- and multi-walled nanotubes on clean silicon
substrates. Since CVD of multi-walled nanotubes on iron occurs at
relatively low temperatures, the metal remains unreacted and
available for nanotube catalysis.  In contrast, the growth of
single-walled nanotubes is performed at higher temperatures where
all of the catalysts have been consumed by the silicide.  In the
case of iron presented in Figure~\ref{Fig2}d, the Fe~2p$_{3/2}$ peak
is only shifted by 0.2~eV at the multi-walled nanotube growth
temperature of 800~$^{\text{o}}$C.  This shift is approximately half
of the binding energy shift that would correspond to a
stoichiometric iron silicide, indicating that catalytic regions of
pure iron are still present which permit high yield growth of
multi-walled nanotubes. By 900~$^{\text{o}}$C the iron silicide
formation is complete, preventing nanotube catalysis and leading to
poor growth yield.

\begin{figure*}[htp]
 \centering
 \includegraphics[width=4.75in]{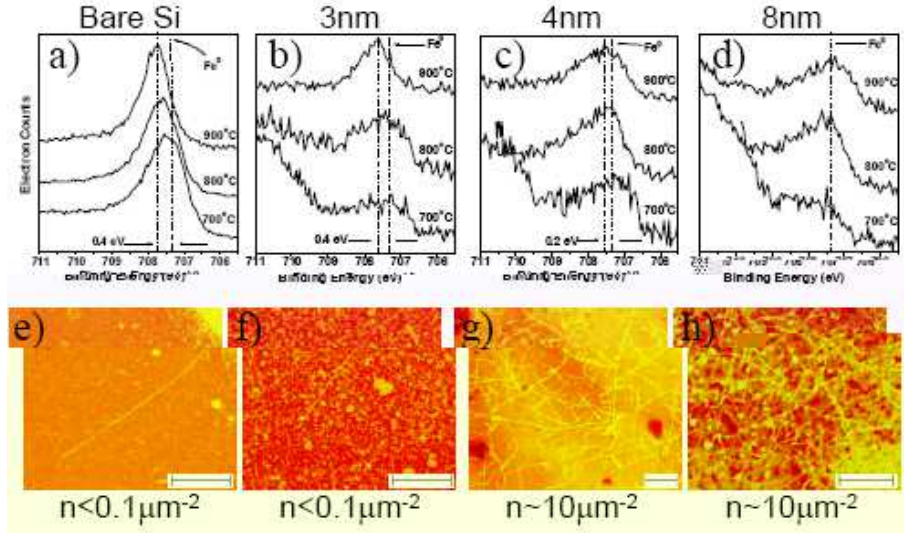}
 \caption{(a-d) Fe~2p$_{3/2}$ core level of iron nitrate on (a) clean silicon, (b) 3~nm SiO$_2$, (c) 4~nm SiO$_2$, and (d) 8~nm SiO$_2$.
 For clean silicon and a 3~nm oxide layer, silicides form by 900~$^{\text{o}}$C, whereas no silicide is detected for the 8~nm oxide.
 On a 4~nm oxide, there is a partial shift in the core level, indicating that the catalyst has not been
 entirely alloyed with the substrate. (e-h) False color SEM images of growths on (e) clean silicon, (f) 3~nm SiO$_2$,
 (g) 4~nm SiO$_2$, and (h) 8~nm SiO$_2$, showing a reduced yield on clean silicon and 3~nm SiO$_2$, but no reduced yield on 4~and
 8~nm oxides. All scale bars are 1~$\mu$m. Images of growth on clean silicon and 3~nm SiO$_2$, (e) and (f), are atypical in that nanotubes
 are absent on most of the substrates. Images (e) and (f) are presented to show the maximal density of nanotubes grown on each
 substrate. The density of grown nanotubes is listed beneath each SEM image.}
 \label{Fig4}
\end{figure*}

Since iron silicide reduces catalytic yield and forms when
iron-based catalysts are used on bare silicon wafers but not on
highly oxidized wafers, it is important to know how to maintain the
iron in a catalytic state without requiring the use of a thick
oxide.  For electronic applications such as FETs, the use of thinner
oxides gives higher gate efficiency due to the increased capacitance
of the gate dielectric.  de los Arcos and co-authors have shown that
iron thin films maintain their catalytic state when they are placed
on top of nitrides (TiN) and oxides (Al$_2$O$_3$ and
TiO$_2$),$^[$\cite{dlArcosAPL,dlArcosCPL,dlArcosCarbon}$^]$ but the
use of silicon oxide would be more compatible with industrial
silicon processing. In order to utilize silicon dioxide, it is
important to determine what is the minimum thickness required to
prevent the formation of a silicide. To determine this, iron
catalyst is deposited onto 3, 4, and 8~nm oxides, and we use the
measured binding energy of the reduced Fe~2p$_{3/2}$ core level
around 707.3~eV to detect any silicide formation
(Figs.~\ref{Fig4}a-d).
If the silicon dioxide layer is thick enough to prevent silicide
formation, the iron core level will remain at 707.3~eV, but if a
silicide forms the peak will shift to higher binding energy, where
707.7~eV is the position of the fully formed iron silicide.  As seen
above for the case of iron or nickel on hydrogen-terminated silicon,
as the substrates are annealed some of the iron oxide is reduced by
the formation of silicon dioxide, even for iron nitrate on an 8~nm
oxide. After this reduction, the metal catalyst could react further
with the substrate to form the silicide.  For catalyst deposited on
the clean silicon (Figure~\ref{Fig4}a) or a 3~nm oxide
(Figure~\ref{Fig4}b), we see that the iron core level shifts to
higher binding energy above 800~$^{\text{o}}$C, indicating the
nucleation of the silicide, followed by complete silicide formation
at 900~$^{\text{o}}$C.  In contrast, heating the catalyst to
900~$^{\text{o}}$C on an 8~nm oxide (Figure~\ref{Fig4}d) does not
lead to silicide formation.  In the case of the 4~nm silicon dioxide
layer (Figure~\ref{Fig4}c), we see a small up-shift of 0.2~eV after
the catalyst has been annealed at 900~$^{\text{o}}$C, suggesting
that some of the iron has been incorporated into a silicide, but
that an appreciable amount of the iron remains in reduced form.
Thus, a 4~nm oxide layer represents the minimum thickness that can
inhibit the formation of the iron silicide during the growth of
single-walled nanotubes at 900~$^{\text{o}}$C.  This value is
similar to a previous estimate of 5-6~nm, where the role of
non-catalytic silicides was conjectured rather than directly
shown.$^[$\cite{CaoAPL}$^]$

\begin{figure}[htp]
 \centering
 \includegraphics[width=2.5in]{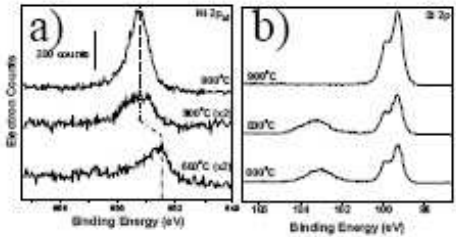}
 \caption{(a) Ni~2p$_{3/2}$ and (b) Si~2p core levels for a nickel film catalyst on native oxide.
 The metallic nickel present at 600~$^{\text{o}}$C has been incorporated into a silicide by 800~$^{\text{o}}$C before the
 desorption of the SiO$_2$ layer.  When the oxide layer is removed, the nickel signal increases
 indicating that there is a higher concentration of nickel beneath the oxide than on top due to diffusion
 of metal through the oxide at lower temperatures.}
 \label{Fig5}
\end{figure}

Having uncovered the minimum barrier oxide thickness in the case of
iron catalyst, it is important to determine what mechanism sets this
limit.  Close observation of the XPS spectra reveals a key signature
of the mechanism of the interfacial reaction:  the silicide is
forming even though a silicon dioxide layer is present on the
surface.  This indicates that either the silicon is diffusing up
through the oxide to form the silicide, or that the metal is
diffusing down.  The direction of diffusion is demonstrated most
clearly in the case of a nickel thin film on a native oxide
substrate (Figure\ref{Fig5}).
At low temperatures the fully reduced nickel film coexists with the
thin SiO$_2$ layer, however by 800~$^{\text{o}}$C the Ni~2p$_{3/2}$
core level is fully shifted due to the formation of the silicide,
while the SiO$_2$ layer remains. After annealing to
900~$^{\text{o}}$C, to desorb the silicon dioxide layer, the spectra
show an increase in intensity from nickel, while the binding energy
remains the same. This increase in intensity is clear evidence that
the nickel silicide was originally beneath the SiO$_2$, where its
XPS intensity was reduced due to the finite electron escape
depth;$^[$\cite{XPSHandbook}$^]$ annealing the sample to remove the
SiO$_2$ eliminates the scattering nickel photoelectrons, leading to
an increase in the nickel intensity.  For the silicide to form
beneath the oxide, the metal catalyst must diffuse through the
oxide. The diffusion mechanism further explains why single-walled
nanotube growth is more inhibited than multi-walled nanotube growth
since the diffusion kinetics are enhanced at the higher
single-walled nanotube growth temperatures.

Given the diffusion mechanism, one can tailor the oxide thickness to
maximize the gate coupling while maintaining the catalyst in an
active state.  Catalyst diffusion through the oxide is determined by
the temperature dependent diffusion constant $D=D_oe^{-\epsilon/kT}$
and the corresponding diffusion length $\lambda\sim\sqrt{Dt}$, where
$\epsilon$ is the activation energy for diffusion and $t$ is the
diffusion time. Using the bulk diffusion constant for iron in
electronic grade silicon dioxide ($D_o\sim10^{-4}$~cm$^2$s$^{-1}$,
$\epsilon=2.8~eV$)$^[$\cite{KononchukAPL}$^]$ one would only expect
the iron to diffuse $\sim1.5$~nm during the temperature ramp to
900~$^{\text{o}}$C, roughly three times shorter than observed in
XPS. The 4~nm diffusion length observed can be explained by a small
reduction of $\sim0.18~eV$ in the activation energy, yielding a
five-fold increase in the effective diffusion constant. An increased
diffusion constant is not surprising for such ultra-thin oxides
since they have higher concentrations of pinhole defects as compared
to thick oxides. Indeed, previous studies of metals on ultra-thin
silicon oxides have shown that pinholes or defects in the oxide
significantly enhances the diffusion
process.$^[$\cite{LiehrPRB,MayerSurfSci,ConfortoPhilMagA,KononchukAPL}$^]$
By determining the critical oxide thickness that prevents silicide
formation for iron catalyst, we can extrapolate to determine the
critical thickness for cobalt and nickel which are the other common
single-walled nanotube catalysts. For cobalt, the bulk diffusion
parameters are $D_o\sim10^{-7}~$cm$^2$s$^{-1}$ and
$\epsilon=1.7~eV$,$^[$\cite{FedorovichSovPhysSS}$^]$ leading to a
diffusion length and critical oxide thickness of $\sim12$~nm at a
900~$^{\text{o}}$C growth temperature.$^[$\cite{OhnoJJAP}$^]$  This
bulk diffusion length is sufficiently thick that high quality oxide
layers can be made to prevent defect-enhanced diffusion.  For
nickel, bulk diffusion
($D_o\sim1.5x10^{-9}~$cm$^2$s$^{-1}$,~$\epsilon=1.6~eV$)$^[$\cite{GhoshJAP}$^]$
would predict a $\sim2.5$~nm critical thickness, which is in the
regime of the enhanced diffusion seen for iron.  If we assume a
proportionate decrease in the activation energy for Ni diffusion
($0.1~eV$), the critical oxide thickness at the 900~$^{\text{o}}$C
growth temperature$^[$\cite{LiuJPCB}$^]$ becomes $\sim4.1$~nm. By
minimizing the concentration of defects in the oxide, one could
reduce these critical oxide thicknesses further. Though the critical
thickness could also be reduced by using a more effective barrier
layer than silicon dioxide, the nanotube-gate coupling will also
change due to changes in the insulator capacitance.

Having determined the minimum oxide thickness to prevent catalyst
diffusion through the oxide layer from XPS, it is important to
ensure that the conclusions are consistent with growth observations.
Surface reactions observed in UHV may be inhibited when the process
is repeated in a CVD reactor where reactant gases can influence
catalyst chemistry.  Since transition-metal silicides are more
stable than transition-metal carbides at nanotube growth
temperatures,$^[$\cite{BackhausActaMeM,PanJACeramS}$^]$ it is
expected that the reactants will not be able to reverse the silicide
formation and nanotube catalysis will be inhibited.  To test these
conclusions, growth was attempted on unannealed pieces of each of
the thin oxides. Figures~\ref{Fig4}e-h show SEM images of the
results of this growth test.  In the case of growth on hydrogen
terminated silicon and 3~nm oxides (Figs.~\ref{Fig4}e,f), nanotubes
are absent on the majority of the substrate. The images shown depict
one of the few locations where nanotubes did grow, and the reported
densities represent an upper limit on the nanotube yield.  As is
readily apparent, there is very little nanotube yield when growth
was performed on hydrogen terminated silicon or a 3~nm oxide.  In
contrast, and consistent with our XPS results, yields on the 4 and
8~nm oxides (figs.~\ref{Fig4}g,h) are significantly improved.

\section{\label{sec:level1}Conclusion}
In summary, we have used x-ray photoelectron spectroscopy to monitor
the formation of silicides that inhibit the growth of carbon
nanotubes on silicon.  Annealing metal catalysts on
hydrogen-terminated silicon or ultra-thin oxide layers leads to the
formation of a non-catalytic silicide around 800~$^{\text{o}}$C (Fe)
and 600~$^{\text{o}}$C (Ni).  By 900~$^{\text{o}}$C, all of the
catalyst has been incorporated into the silicide.  In order to
prevent silicide formation, the oxide substrate must act as an
effective barrier to the diffusion of metal catalyst at the high
single-walled nanotube growth temperature.  In the case of iron, our
XPS and growth experiments have shown that oxide layers that are
4~nm or greater are able to inhibit silicide formation at
900~$^{\text{o}}$C, leading to a high nanotube growth yield.  By
limiting catalyst diffusion through the oxide layer, either with
reduced growth temperatures or reduced diffusivity in the oxide,
single-walled nanotubes can be grown with high yields on ultra-thin
SiO$_2$ layers.

\section{\label{sec:level1}Experimental}
The substrates used in this study consist of highly doped silicon
wafers capped with thermally grown SiO$_2$ (100, 8, 4, or 3~nm), a
native ($\sim1$~nm) oxide, or hydrogen termination.  All oxide
thicknesses are measured using multi-wavelength ellipsometry and
confirmed by direct measurements using atomic force microscopy.  The
oxide thickness uniformity across the substrates is better than
1~nm.  To obtain a clean silicon wafer, a hydrofluoric acid etch is
used to remove the native oxide and hydrogen-terminate the silicon.
Immediately after hydrogen termination, the catalyst is deposited
and the sample is introduced into the vacuum chamber within five
minutes in order to limit the formation of a native oxide layer. The
quality of the hydrogen termination is confirmed by the low
intensity of the oxidized silicon peak in the XPS spectra.

Although there are a large number of catalysts available, many
require the incorporation of an insulating matrix, such as alumina,
which can influence the chemistry of the catalyst.  For example,
annealing a thin film of iron on an alumina buffer layer was shown
to lead to an oxidation of the iron to
Fe$_2$O$_3$.$^[$\cite{dlArcosCPL,dlArcosCarbon}$^]$  In order to
avoid such interactions, we use catalysts that do not require a
matrix: iron nitrate and thermally evaporated nickel thin films. The
iron nitrate catalyst is prepared by dissolving 50~mg
iron~(III)~nitrate nonahydrate (Fe(NO$_3$)$_3~\cdot~$9H$_2$O) in
100~ml isopropanol and stirring for two
minutes.$^[$\cite{HafnerJPCB}$^]$ The substrates are then soaked in
the solution for one minute and the sample is dried with flowing
nitrogen, forming a thin layer of catalyst over the entire
substrate.  Nickel catalyst is deposited onto the substrates by
electron beam evaporation, resulting in a 7~{\AA} film which
decomposes at the growth temperature to form nanoparticles. Chemical
vapor deposition tests are performed in an atmospheric pressure
reactor.  The substrates are brought to the 900~$^{\text{o}}$C
growth temperature under argon flow, after which methane (400~sccm)
and hydrogen (20~sccm) are used for the 10 minute nanotube growth.
The samples are then cooled to room temperature under argon before
being removed from the reactor.

XPS is performed in an ion-pumped ultra-high vacuum (UHV) chamber
with a working pressure of better than 2x10$^{-9}$~torr, using
monochromatic Al$_{K_{\alpha}}$ (1486.6~eV) x-rays at a resolution
of $\sim0.2$~eV. Sample heating is performed in UHV by passing
current through the silicon substrate and the temperature is
measured using an optical pyrometer.  After introduction into the
UHV chamber, the samples are outgassed at $\sim$100~$^{\text{o}}$C
for 10 minutes to remove any residual contaminants.  In order to
mimic the kinetics of the temperature ramp used in a typical CVD
growth, the sample is slowly heated from room temperature to the
anneal temperature over 5 minutes.  After reaching the desired
temperature, substrate heating is stopped and the sample is allowed
to cool to room temperature. During the anneal, the base pressure in
the chamber increases, preventing the acquisition of spectra at the
anneal temperature. All energies are corrected to the Si~2p$_{3/2}$
core level at 99.3~eV for clean silicon and 103.3~eV for
SiO$_2$$^[$\cite{XPSHandbook}$^]$ to correct for charging effects on
the oxide substrates.  Quantitative XPS fitting is obtained by
fitting raw data to Voigt functions after a Shirley background
correction.$^[$\cite{ShirleyPRB}$^]$

\begin{acknowledgments}
The authors acknowledge financial support from the NSF MRSEC program
under DMR-0520527, the NSF CAREER program under DMR-0094063, the NSF
NSEC program under DMR-0425880, and DMR-0210806.
\end{acknowledgments}


\newpage
\begin{thebibliography}{51}
\expandafter\ifx\csname natexlab\endcsname\relax\def\natexlab#1{#1}\fi
\expandafter\ifx\csname bibnamefont\endcsname\relax
  \def\bibnamefont#1{#1}\fi
\expandafter\ifx\csname bibfnamefont\endcsname\relax
  \def\bibfnamefont#1{#1}\fi
\expandafter\ifx\csname citenamefont\endcsname\relax
  \def\citenamefont#1{#1}\fi
\expandafter\ifx\csname url\endcsname\relax
  \def\url#1{\texttt{#1}}\fi
\expandafter\ifx\csname urlprefix\endcsname\relax\def\urlprefix{URL }\fi
\providecommand{\bibinfo}[2]{#2}
\providecommand{\eprint}[2][]{\url{#2}}

\bibitem[{\citenamefont{Saito et~al.}(1998)\citenamefont{Saito, Dresselhaus,
  and Dresselhaus}}]{Saitobook}
\bibinfo{author}{\bibfnamefont{R.}~\bibnamefont{Saito}},
  \bibinfo{author}{\bibfnamefont{G.}~\bibnamefont{Dresselhaus}},
  \bibnamefont{and} \bibinfo{author}{\bibfnamefont{M.~S.}
  \bibnamefont{Dresselhaus}}, \emph{\bibinfo{title}{Physical Properties of
  Carbon Nanotubes}} (\bibinfo{publisher}{Imperial College Press},
  \bibinfo{address}{London}, \bibinfo{year}{1998}).

\bibitem[{\citenamefont{Martel et~al.}(1998)\citenamefont{Martel, Schmidt,
  Shea, Hertel, and Avouris}}]{MartelAPL}
\bibinfo{author}{\bibfnamefont{R.}~\bibnamefont{Martel}},
  \bibinfo{author}{\bibfnamefont{T.}~\bibnamefont{Schmidt}},
  \bibinfo{author}{\bibfnamefont{H.~R.} \bibnamefont{Shea}},
  \bibinfo{author}{\bibfnamefont{T.}~\bibnamefont{Hertel}}, \bibnamefont{and}
  \bibinfo{author}{\bibfnamefont{P.}~\bibnamefont{Avouris}},
  \bibinfo{journal}{Applied Physics Letters} \textbf{\bibinfo{volume}{73}},
  \bibinfo{pages}{2447} (\bibinfo{year}{1998}).

\bibitem[{\citenamefont{Tans et~al.}(1998)\citenamefont{Tans, Verschueren, and
  Dekker}}]{TansNature}
\bibinfo{author}{\bibfnamefont{S.~J.} \bibnamefont{Tans}},
  \bibinfo{author}{\bibfnamefont{A.~R.~M.} \bibnamefont{Verschueren}},
  \bibnamefont{and} \bibinfo{author}{\bibfnamefont{C.}~\bibnamefont{Dekker}},
  \bibinfo{journal}{Nature} \textbf{\bibinfo{volume}{393}}, \bibinfo{pages}{49}
  (\bibinfo{year}{1998}).

\bibitem[{\citenamefont{Chopra et~al.}(2003)\citenamefont{Chopra, McGuire,
  Gothard, Rao, and Pham}}]{ChopraAPL}
\bibinfo{author}{\bibfnamefont{S.}~\bibnamefont{Chopra}},
  \bibinfo{author}{\bibfnamefont{K.}~\bibnamefont{McGuire}},
  \bibinfo{author}{\bibfnamefont{N.}~\bibnamefont{Gothard}},
  \bibinfo{author}{\bibfnamefont{A.~M.} \bibnamefont{Rao}}, \bibnamefont{and}
  \bibinfo{author}{\bibfnamefont{A.}~\bibnamefont{Pham}},
  \bibinfo{journal}{Applied Physics Letters} \textbf{\bibinfo{volume}{83}},
  \bibinfo{pages}{2280} (\bibinfo{year}{2003}).

\bibitem[{\citenamefont{Kong et~al.}(2000)\citenamefont{Kong, Franklin, Zhou,
  Chapline, Peng, Cho, and Dai}}]{KongScience}
\bibinfo{author}{\bibfnamefont{J.}~\bibnamefont{Kong}},
  \bibinfo{author}{\bibfnamefont{N.~R.} \bibnamefont{Franklin}},
  \bibinfo{author}{\bibfnamefont{C.}~\bibnamefont{Zhou}},
  \bibinfo{author}{\bibfnamefont{M.~G.} \bibnamefont{Chapline}},
  \bibinfo{author}{\bibfnamefont{S.}~\bibnamefont{Peng}},
  \bibinfo{author}{\bibfnamefont{K.}~\bibnamefont{Cho}}, \bibnamefont{and}
  \bibinfo{author}{\bibfnamefont{H.}~\bibnamefont{Dai}},
  \bibinfo{journal}{Science} \textbf{\bibinfo{volume}{287}},
  \bibinfo{pages}{622} (\bibinfo{year}{2000}).

\bibitem[{\citenamefont{Lee et~al.}(2004)\citenamefont{Lee, Baker, Marcus,
  Yang, Eriksson, and Hamers}}]{LeeNanoLett}
\bibinfo{author}{\bibfnamefont{C.~S.} \bibnamefont{Lee}},
  \bibinfo{author}{\bibfnamefont{S.~E.} \bibnamefont{Baker}},
  \bibinfo{author}{\bibfnamefont{M.~S.} \bibnamefont{Marcus}},
  \bibinfo{author}{\bibfnamefont{W.}~\bibnamefont{Yang}},
  \bibinfo{author}{\bibfnamefont{M.~A.} \bibnamefont{Eriksson}},
  \bibnamefont{and} \bibinfo{author}{\bibfnamefont{R.~J.}
  \bibnamefont{Hamers}}, \bibinfo{journal}{Nano Letters}
  \textbf{\bibinfo{volume}{4}}, \bibinfo{pages}{1713} (\bibinfo{year}{2004}).

\bibitem[{\citenamefont{Dai et~al.}(1999)\citenamefont{Dai, Kong, Zhou,
  Franklin, Tombler, Cassell, Fan, and Chapline}}]{DaiJPCB}
\bibinfo{author}{\bibfnamefont{H.}~\bibnamefont{Dai}},
  \bibinfo{author}{\bibfnamefont{J.}~\bibnamefont{Kong}},
  \bibinfo{author}{\bibfnamefont{C.}~\bibnamefont{Zhou}},
  \bibinfo{author}{\bibfnamefont{N.}~\bibnamefont{Franklin}},
  \bibinfo{author}{\bibfnamefont{T.}~\bibnamefont{Tombler}},
  \bibinfo{author}{\bibfnamefont{A.}~\bibnamefont{Cassell}},
  \bibinfo{author}{\bibfnamefont{S.}~\bibnamefont{Fan}}, \bibnamefont{and}
  \bibinfo{author}{\bibfnamefont{M.}~\bibnamefont{Chapline}},
  \bibinfo{journal}{Journal of Physical Chemistry B}
  \textbf{\bibinfo{volume}{103}}, \bibinfo{pages}{11246}
  (\bibinfo{year}{1999}).

\bibitem[{\citenamefont{Kong et~al.}(1998)\citenamefont{Kong, Cassell, and
  Dai}}]{KongCPL}
\bibinfo{author}{\bibfnamefont{J.}~\bibnamefont{Kong}},
  \bibinfo{author}{\bibfnamefont{A.~M.} \bibnamefont{Cassell}},
  \bibnamefont{and} \bibinfo{author}{\bibfnamefont{H.}~\bibnamefont{Dai}},
  \bibinfo{journal}{Chemical Physics Letters} \textbf{\bibinfo{volume}{292}},
  \bibinfo{pages}{567} (\bibinfo{year}{1998}).

\bibitem[{\citenamefont{Moisala et~al.}(2003)\citenamefont{Moisala, Nasibulin,
  and Kauppinen}}]{MoisalaJPCondMatt}
\bibinfo{author}{\bibfnamefont{A.}~\bibnamefont{Moisala}},
  \bibinfo{author}{\bibfnamefont{A.}~\bibnamefont{Nasibulin}},
  \bibnamefont{and} \bibinfo{author}{\bibfnamefont{E.~I.}
  \bibnamefont{Kauppinen}}, \bibinfo{journal}{Journal of Physics: Condensed
  Matter} \textbf{\bibinfo{volume}{15}}, \bibinfo{pages}{S3011}
  (\bibinfo{year}{2003}).

\bibitem[{\citenamefont{Baker et~al.}(1982)\citenamefont{Baker, Alonzo,
  Dumesic, and Yates}}]{BakerJCatal}
\bibinfo{author}{\bibfnamefont{R.~T.~K.} \bibnamefont{Baker}},
  \bibinfo{author}{\bibfnamefont{J.~R.} \bibnamefont{Alonzo}},
  \bibinfo{author}{\bibfnamefont{J.~A.} \bibnamefont{Dumesic}},
  \bibnamefont{and} \bibinfo{author}{\bibfnamefont{D.~J.~C.}
  \bibnamefont{Yates}}, \bibinfo{journal}{Journal of Catalysis}
  \textbf{\bibinfo{volume}{77}}, \bibinfo{pages}{74} (\bibinfo{year}{1982}).

\bibitem[{\citenamefont{de~los Arcos
  et~al.}(2004{\natexlab{a}})\citenamefont{de~los Arcos, Garnier, Seo,
  Oelhafen, Thommen, and Mathys}}]{dlArcosJPCB}
\bibinfo{author}{\bibfnamefont{T.}~\bibnamefont{de~los Arcos}},
  \bibinfo{author}{\bibfnamefont{M.~G.} \bibnamefont{Garnier}},
  \bibinfo{author}{\bibfnamefont{J.~W.} \bibnamefont{Seo}},
  \bibinfo{author}{\bibfnamefont{P.}~\bibnamefont{Oelhafen}},
  \bibinfo{author}{\bibfnamefont{V.}~\bibnamefont{Thommen}}, \bibnamefont{and}
  \bibinfo{author}{\bibfnamefont{D.}~\bibnamefont{Mathys}},
  \bibinfo{journal}{Journal of Physical Chemistry B}
  \textbf{\bibinfo{volume}{108}}, \bibinfo{pages}{7728}
  (\bibinfo{year}{2004}{\natexlab{a}}).

\bibitem[{\citenamefont{de~los Arcos et~al.}(2002)\citenamefont{de~los Arcos,
  Vonau, Garnier, Thommen, Boyen, Oelhafen, Duggelin, Mathis, and
  Guggenheim}}]{dlArcosAPL}
\bibinfo{author}{\bibfnamefont{T.}~\bibnamefont{de~los Arcos}},
  \bibinfo{author}{\bibfnamefont{F.}~\bibnamefont{Vonau}},
  \bibinfo{author}{\bibfnamefont{M.~G.} \bibnamefont{Garnier}},
  \bibinfo{author}{\bibfnamefont{V.}~\bibnamefont{Thommen}},
  \bibinfo{author}{\bibfnamefont{H.-G.} \bibnamefont{Boyen}},
  \bibinfo{author}{\bibfnamefont{P.}~\bibnamefont{Oelhafen}},
  \bibinfo{author}{\bibfnamefont{M.}~\bibnamefont{Duggelin}},
  \bibinfo{author}{\bibfnamefont{D.}~\bibnamefont{Mathis}}, \bibnamefont{and}
  \bibinfo{author}{\bibfnamefont{R.}~\bibnamefont{Guggenheim}},
  \bibinfo{journal}{Applied Physics Letters} \textbf{\bibinfo{volume}{80}},
  \bibinfo{pages}{2383} (\bibinfo{year}{2002}).

\bibitem[{\citenamefont{Klinke et~al.}(2004)\citenamefont{Klinke, Bonard, and
  Kern}}]{KlinkeJPCB}
\bibinfo{author}{\bibfnamefont{C.}~\bibnamefont{Klinke}},
  \bibinfo{author}{\bibfnamefont{J.~M.} \bibnamefont{Bonard}},
  \bibnamefont{and} \bibinfo{author}{\bibfnamefont{K.}~\bibnamefont{Kern}},
  \bibinfo{journal}{Journal of Physical Chemistry B}
  \textbf{\bibinfo{volume}{108}}, \bibinfo{pages}{11357}
  (\bibinfo{year}{2004}).

\bibitem[{\citenamefont{Konya et~al.}(2001)\citenamefont{Konya, Kiss, Oszko,
  Siska, and Kiricsi}}]{KonyaPCCP}
\bibinfo{author}{\bibfnamefont{Z.}~\bibnamefont{Konya}},
  \bibinfo{author}{\bibfnamefont{J.}~\bibnamefont{Kiss}},
  \bibinfo{author}{\bibfnamefont{A.}~\bibnamefont{Oszko}},
  \bibinfo{author}{\bibfnamefont{A.}~\bibnamefont{Siska}}, \bibnamefont{and}
  \bibinfo{author}{\bibfnamefont{I.}~\bibnamefont{Kiricsi}},
  \bibinfo{journal}{Physical Chemistry Chemical Physics}
  \textbf{\bibinfo{volume}{3}}, \bibinfo{pages}{155} (\bibinfo{year}{2001}).

\bibitem[{\citenamefont{Emmenegger et~al.}(2003)\citenamefont{Emmenegger,
  Bonard, Mauron, Sudan, Lepora, Grobety, Zuttel, and
  Schlapbach}}]{EmmeneggerCarbon}
\bibinfo{author}{\bibfnamefont{C.}~\bibnamefont{Emmenegger}},
  \bibinfo{author}{\bibfnamefont{J.~M.} \bibnamefont{Bonard}},
  \bibinfo{author}{\bibfnamefont{P.}~\bibnamefont{Mauron}},
  \bibinfo{author}{\bibfnamefont{P.}~\bibnamefont{Sudan}},
  \bibinfo{author}{\bibfnamefont{A.}~\bibnamefont{Lepora}},
  \bibinfo{author}{\bibfnamefont{B.}~\bibnamefont{Grobety}},
  \bibinfo{author}{\bibfnamefont{A.}~\bibnamefont{Zuttel}}, \bibnamefont{and}
  \bibinfo{author}{\bibfnamefont{L.}~\bibnamefont{Schlapbach}},
  \bibinfo{journal}{Carbon} \textbf{\bibinfo{volume}{41}}, \bibinfo{pages}{539}
  (\bibinfo{year}{2003}).

\bibitem[{\citenamefont{Prabhakaran et~al.}(2003)\citenamefont{Prabhakaran,
  Watanabe, Homma, Ogino, Wei, Ajayan, Shafi, Ulman, Heun, Locatelli
  et~al.}}]{PrabhakaranLang}
\bibinfo{author}{\bibfnamefont{K.}~\bibnamefont{Prabhakaran}},
  \bibinfo{author}{\bibfnamefont{Y.}~\bibnamefont{Watanabe}},
  \bibinfo{author}{\bibfnamefont{Y.}~\bibnamefont{Homma}},
  \bibinfo{author}{\bibfnamefont{T.}~\bibnamefont{Ogino}},
  \bibinfo{author}{\bibfnamefont{B.~Q.} \bibnamefont{Wei}},
  \bibinfo{author}{\bibfnamefont{P.~M.} \bibnamefont{Ajayan}},
  \bibinfo{author}{\bibfnamefont{K.}~\bibnamefont{Shafi}},
  \bibinfo{author}{\bibfnamefont{A.}~\bibnamefont{Ulman}},
  \bibinfo{author}{\bibfnamefont{S.}~\bibnamefont{Heun}},
  \bibinfo{author}{\bibfnamefont{A.}~\bibnamefont{Locatelli}},
  \bibnamefont{et~al.}, \bibinfo{journal}{Langmuir}
  \textbf{\bibinfo{volume}{19}}, \bibinfo{pages}{10629} (\bibinfo{year}{2003}).

\bibitem[{\citenamefont{de~los Arcos et~al.}(2003)\citenamefont{de~los Arcos,
  Wu, and Oelhafen}}]{dlArcosCPL}
\bibinfo{author}{\bibfnamefont{T.}~\bibnamefont{de~los Arcos}},
  \bibinfo{author}{\bibfnamefont{Z.~M.} \bibnamefont{Wu}}, \bibnamefont{and}
  \bibinfo{author}{\bibfnamefont{P.}~\bibnamefont{Oelhafen}},
  \bibinfo{journal}{Chemical Physics Letters} \textbf{\bibinfo{volume}{380}},
  \bibinfo{pages}{419} (\bibinfo{year}{2003}).

\bibitem[{\citenamefont{Herrera and Resasco}(2004)}]{HerreraJCatal}
\bibinfo{author}{\bibfnamefont{J.~E.} \bibnamefont{Herrera}} \bibnamefont{and}
  \bibinfo{author}{\bibfnamefont{D.~E.} \bibnamefont{Resasco}},
  \bibinfo{journal}{Journal of Catalysis} \textbf{\bibinfo{volume}{221}},
  \bibinfo{pages}{354} (\bibinfo{year}{2004}).

\bibitem[{\citenamefont{Nishimura et~al.}(2004)\citenamefont{Nishimura,
  Okazaki, Pan, and Nakayama}}]{NishimuraJJAP}
\bibinfo{author}{\bibfnamefont{K.}~\bibnamefont{Nishimura}},
  \bibinfo{author}{\bibfnamefont{N.}~\bibnamefont{Okazaki}},
  \bibinfo{author}{\bibfnamefont{L.~J.} \bibnamefont{Pan}}, \bibnamefont{and}
  \bibinfo{author}{\bibfnamefont{Y.}~\bibnamefont{Nakayama}},
  \bibinfo{journal}{Japanese Journal of Applied Physics}
  \textbf{\bibinfo{volume}{43}}, \bibinfo{pages}{L471} (\bibinfo{year}{2004}).

\bibitem[{\citenamefont{Yang et~al.}(2005)\citenamefont{Yang, Marcus, Keppel,
  Zhang, Li, Larson, Savage, Simmons, Castellini, Eriksson et~al.}}]{YangAPL}
\bibinfo{author}{\bibfnamefont{B.}~\bibnamefont{Yang}},
  \bibinfo{author}{\bibfnamefont{M.~S.} \bibnamefont{Marcus}},
  \bibinfo{author}{\bibfnamefont{D.~G.} \bibnamefont{Keppel}},
  \bibinfo{author}{\bibfnamefont{P.~P.} \bibnamefont{Zhang}},
  \bibinfo{author}{\bibfnamefont{Z.~W.} \bibnamefont{Li}},
  \bibinfo{author}{\bibfnamefont{B.~J.} \bibnamefont{Larson}},
  \bibinfo{author}{\bibfnamefont{D.~E.} \bibnamefont{Savage}},
  \bibinfo{author}{\bibfnamefont{J.~M.} \bibnamefont{Simmons}},
  \bibinfo{author}{\bibfnamefont{O.~M.} \bibnamefont{Castellini}},
  \bibinfo{author}{\bibfnamefont{M.~A.} \bibnamefont{Eriksson}},
  \bibnamefont{et~al.}, \bibinfo{journal}{Applied Physics Letters}
  \textbf{\bibinfo{volume}{86}}, \bibinfo{pages}{263107}
  (\bibinfo{year}{2005}).

\bibitem[{\citenamefont{Jung et~al.}(2003)\citenamefont{Jung, Wei, Vajtai,
  Ajayan, Homma, Prabhakaran, and Ogino}}]{JungNanoLett}
\bibinfo{author}{\bibfnamefont{Y.}~\bibnamefont{Jung}},
  \bibinfo{author}{\bibfnamefont{B.~Q.} \bibnamefont{Wei}},
  \bibinfo{author}{\bibfnamefont{R.}~\bibnamefont{Vajtai}},
  \bibinfo{author}{\bibfnamefont{P.~M.} \bibnamefont{Ajayan}},
  \bibinfo{author}{\bibfnamefont{Y.}~\bibnamefont{Homma}},
  \bibinfo{author}{\bibfnamefont{K.}~\bibnamefont{Prabhakaran}},
  \bibnamefont{and} \bibinfo{author}{\bibfnamefont{T.}~\bibnamefont{Ogino}},
  \bibinfo{journal}{Nano Letters} \textbf{\bibinfo{volume}{3}},
  \bibinfo{pages}{561} (\bibinfo{year}{2003}).

\bibitem[{\citenamefont{Maeda et~al.}(2004)\citenamefont{Maeda, Laffosse,
  Watanabe, Suzuki, Homma, Suzuki, Kitada, Ogiwara, Tanaka, Kimura
  et~al.}}]{MaedaPhysicaE}
\bibinfo{author}{\bibfnamefont{F.}~\bibnamefont{Maeda}},
  \bibinfo{author}{\bibfnamefont{E.}~\bibnamefont{Laffosse}},
  \bibinfo{author}{\bibfnamefont{Y.}~\bibnamefont{Watanabe}},
  \bibinfo{author}{\bibfnamefont{S.}~\bibnamefont{Suzuki}},
  \bibinfo{author}{\bibfnamefont{Y.}~\bibnamefont{Homma}},
  \bibinfo{author}{\bibfnamefont{M.}~\bibnamefont{Suzuki}},
  \bibinfo{author}{\bibfnamefont{T.}~\bibnamefont{Kitada}},
  \bibinfo{author}{\bibfnamefont{T.}~\bibnamefont{Ogiwara}},
  \bibinfo{author}{\bibfnamefont{A.}~\bibnamefont{Tanaka}},
  \bibinfo{author}{\bibfnamefont{M.}~\bibnamefont{Kimura}},
  \bibnamefont{et~al.}, \bibinfo{journal}{Physica E}
  \textbf{\bibinfo{volume}{24}}, \bibinfo{pages}{19} (\bibinfo{year}{2004}).

\bibitem[{\citenamefont{Sohn et~al.}(2001)\citenamefont{Sohn, Choi, Lee, and
  Seong}}]{SohnAPL}
\bibinfo{author}{\bibfnamefont{J.~I.} \bibnamefont{Sohn}},
  \bibinfo{author}{\bibfnamefont{C.~J.} \bibnamefont{Choi}},
  \bibinfo{author}{\bibfnamefont{S.}~\bibnamefont{Lee}}, \bibnamefont{and}
  \bibinfo{author}{\bibfnamefont{T.~Y.} \bibnamefont{Seong}},
  \bibinfo{journal}{Applied Physics Letters} \textbf{\bibinfo{volume}{78}},
  \bibinfo{pages}{3130} (\bibinfo{year}{2001}).

\bibitem[{\citenamefont{Cao et~al.}(2004)\citenamefont{Cao, Ajayan, Ramanath,
  Baskaran, and Turner}}]{CaoAPL}
\bibinfo{author}{\bibfnamefont{A.}~\bibnamefont{Cao}},
  \bibinfo{author}{\bibfnamefont{P.~M.} \bibnamefont{Ajayan}},
  \bibinfo{author}{\bibfnamefont{G.}~\bibnamefont{Ramanath}},
  \bibinfo{author}{\bibfnamefont{R.}~\bibnamefont{Baskaran}}, \bibnamefont{and}
  \bibinfo{author}{\bibfnamefont{K.}~\bibnamefont{Turner}},
  \bibinfo{journal}{Applied Physics Letters} \textbf{\bibinfo{volume}{84}},
  \bibinfo{pages}{109} (\bibinfo{year}{2004}).

\bibitem[{\citenamefont{Cheung et~al.}(2000)\citenamefont{Cheung, Hafner, and
  Lieber}}]{CheungPNAS}
\bibinfo{author}{\bibfnamefont{C.~L.} \bibnamefont{Cheung}},
  \bibinfo{author}{\bibfnamefont{J.~H.} \bibnamefont{Hafner}},
  \bibnamefont{and} \bibinfo{author}{\bibfnamefont{C.~M.}
  \bibnamefont{Lieber}}, \bibinfo{journal}{Proceedings of the National Academy
  of Sciences} \textbf{\bibinfo{volume}{97}}, \bibinfo{pages}{3809}
  (\bibinfo{year}{2000}).

\bibitem[{\citenamefont{Kondo et~al.}(2005)\citenamefont{Kondo, Sato, Kawabata,
  and Awano}}]{KondoJJAP}
\bibinfo{author}{\bibfnamefont{D.}~\bibnamefont{Kondo}},
  \bibinfo{author}{\bibfnamefont{S.}~\bibnamefont{Sato}},
  \bibinfo{author}{\bibfnamefont{A.}~\bibnamefont{Kawabata}}, \bibnamefont{and}
  \bibinfo{author}{\bibfnamefont{Y.}~\bibnamefont{Awano}},
  \bibinfo{journal}{Japanese Journal Of Applied Physics}
  \textbf{\bibinfo{volume}{44}}, \bibinfo{pages}{5292} (\bibinfo{year}{2005}).

\bibitem[{\citenamefont{Wong et~al.}(2005)\citenamefont{Wong, Lee, and
  Lee}}]{WongAPL}
\bibinfo{author}{\bibfnamefont{W.~K.} \bibnamefont{Wong}},
  \bibinfo{author}{\bibfnamefont{C.~S.} \bibnamefont{Lee}}, \bibnamefont{and}
  \bibinfo{author}{\bibfnamefont{S.~T.} \bibnamefont{Lee}},
  \bibinfo{journal}{Journal of Applied Physics} \textbf{\bibinfo{volume}{97}},
  \bibinfo{pages}{084307} (\bibinfo{year}{2005}).

\bibitem[{\citenamefont{McIntyre and Zetaruk}(1977)}]{McIntyreAnalChem}
\bibinfo{author}{\bibfnamefont{N.~S.} \bibnamefont{McIntyre}} \bibnamefont{and}
  \bibinfo{author}{\bibfnamefont{D.~G.} \bibnamefont{Zetaruk}},
  \bibinfo{journal}{Analytical Chemistry} \textbf{\bibinfo{volume}{49}},
  \bibinfo{pages}{1521} (\bibinfo{year}{1977}).

\bibitem[{\citenamefont{Lin et~al.}(1997)\citenamefont{Lin, Seshadri, and
  Kelber}}]{LinApplSurfSci}
\bibinfo{author}{\bibfnamefont{T.~C.} \bibnamefont{Lin}},
  \bibinfo{author}{\bibfnamefont{G.}~\bibnamefont{Seshadri}}, \bibnamefont{and}
  \bibinfo{author}{\bibfnamefont{J.~A.} \bibnamefont{Kelber}},
  \bibinfo{journal}{Applied Surface Science} \textbf{\bibinfo{volume}{119}},
  \bibinfo{pages}{83} (\bibinfo{year}{1997}).

\bibitem[{\citenamefont{Grosvenor et~al.}(2004)\citenamefont{Grosvenor, Kobe,
  Biesinger, and McIntyre}}]{GrosvenorSurfInterfaceAnal}
\bibinfo{author}{\bibfnamefont{A.~P.} \bibnamefont{Grosvenor}},
  \bibinfo{author}{\bibfnamefont{B.~A.} \bibnamefont{Kobe}},
  \bibinfo{author}{\bibfnamefont{M.~C.} \bibnamefont{Biesinger}},
  \bibnamefont{and} \bibinfo{author}{\bibfnamefont{N.~S.}
  \bibnamefont{McIntyre}}, \bibinfo{journal}{Surface And Interface Analysis}
  \textbf{\bibinfo{volume}{36}}, \bibinfo{pages}{1564} (\bibinfo{year}{2004}).

\bibitem[{\citenamefont{Chase}(1998)}]{NIST-JANAF}
\bibinfo{author}{\bibfnamefont{M.~W.} \bibnamefont{Chase}},
  \bibinfo{journal}{Journal of Physical and Chemical Reference Data} p.
  \bibinfo{pages}{Monograph 9} (\bibinfo{year}{1998}).

\bibitem[{\citenamefont{Orlowski et~al.}(2004)\citenamefont{Orlowski, Kowalski,
  Fronc, Zuberek, Mickevicius, Mirabella, and Ghijsen}}]{OrlowskiJAlloysCmpd}
\bibinfo{author}{\bibfnamefont{B.~A.} \bibnamefont{Orlowski}},
  \bibinfo{author}{\bibfnamefont{B.~J.} \bibnamefont{Kowalski}},
  \bibinfo{author}{\bibfnamefont{K.}~\bibnamefont{Fronc}},
  \bibinfo{author}{\bibfnamefont{R.}~\bibnamefont{Zuberek}},
  \bibinfo{author}{\bibfnamefont{S.}~\bibnamefont{Mickevicius}},
  \bibinfo{author}{\bibfnamefont{F.}~\bibnamefont{Mirabella}},
  \bibnamefont{and} \bibinfo{author}{\bibfnamefont{J.}~\bibnamefont{Ghijsen}},
  \bibinfo{journal}{Journal of Alloys and Compounds}
  \textbf{\bibinfo{volume}{362}}, \bibinfo{pages}{202} (\bibinfo{year}{2004}).

\bibitem[{\citenamefont{Homma et~al.}(2003)\citenamefont{Homma, Kobayashi,
  Ogino, Takagi, Ito, Jung, and Ajayan}}]{HommaJPCB}
\bibinfo{author}{\bibfnamefont{Y.}~\bibnamefont{Homma}},
  \bibinfo{author}{\bibfnamefont{Y.}~\bibnamefont{Kobayashi}},
  \bibinfo{author}{\bibfnamefont{T.}~\bibnamefont{Ogino}},
  \bibinfo{author}{\bibfnamefont{D.}~\bibnamefont{Takagi}},
  \bibinfo{author}{\bibfnamefont{R.}~\bibnamefont{Ito}},
  \bibinfo{author}{\bibfnamefont{Y.}~\bibnamefont{Jung}}, \bibnamefont{and}
  \bibinfo{author}{\bibfnamefont{P.}~\bibnamefont{Ajayan}},
  \bibinfo{journal}{Journal of Physical Chemistry B}
  \textbf{\bibinfo{volume}{107}}, \bibinfo{pages}{12161}
  (\bibinfo{year}{2003}).

\bibitem[{\citenamefont{Ruhrnschopf et~al.}(1996)\citenamefont{Ruhrnschopf,
  Borgmann, and Wedler}}]{RuhrnschopfTSF}
\bibinfo{author}{\bibfnamefont{K.}~\bibnamefont{Ruhrnschopf}},
  \bibinfo{author}{\bibfnamefont{D.}~\bibnamefont{Borgmann}}, \bibnamefont{and}
  \bibinfo{author}{\bibfnamefont{G.}~\bibnamefont{Wedler}},
  \bibinfo{journal}{Thin Solid Films} \textbf{\bibinfo{volume}{280}},
  \bibinfo{pages}{171} (\bibinfo{year}{1996}).

\bibitem[{\citenamefont{Zhu et~al.}(1986)\citenamefont{Zhu, Iwasaki, Williams,
  and Park}}]{ZhuJAP}
\bibinfo{author}{\bibfnamefont{Q.~G.} \bibnamefont{Zhu}},
  \bibinfo{author}{\bibfnamefont{H.}~\bibnamefont{Iwasaki}},
  \bibinfo{author}{\bibfnamefont{E.~D.} \bibnamefont{Williams}},
  \bibnamefont{and} \bibinfo{author}{\bibfnamefont{R.~L.} \bibnamefont{Park}},
  \bibinfo{journal}{Journal of Applied Physics} \textbf{\bibinfo{volume}{60}},
  \bibinfo{pages}{2629} (\bibinfo{year}{1986}).

\bibitem[{\citenamefont{Cheung et~al.}(1981)\citenamefont{Cheung, Grunthaner,
  Grunthaner, Mayer, and Ullrich}}]{CheungJVST}
\bibinfo{author}{\bibfnamefont{N.~W.} \bibnamefont{Cheung}},
  \bibinfo{author}{\bibfnamefont{P.~J.} \bibnamefont{Grunthaner}},
  \bibinfo{author}{\bibfnamefont{F.~J.} \bibnamefont{Grunthaner}},
  \bibinfo{author}{\bibfnamefont{J.~W.} \bibnamefont{Mayer}}, \bibnamefont{and}
  \bibinfo{author}{\bibfnamefont{B.~M.} \bibnamefont{Ullrich}},
  \bibinfo{journal}{Journal of Vacuum Science and Technology}
  \textbf{\bibinfo{volume}{18}}, \bibinfo{pages}{917} (\bibinfo{year}{1981}).

\bibitem[{\citenamefont{Nguyen and Cinti}(1983)}]{NguyenPhysScripta}
\bibinfo{author}{\bibfnamefont{T.~T.~A.} \bibnamefont{Nguyen}}
  \bibnamefont{and} \bibinfo{author}{\bibfnamefont{R.}~\bibnamefont{Cinti}},
  \bibinfo{journal}{Physica Scripta} \textbf{\bibinfo{volume}{T4}},
  \bibinfo{pages}{176} (\bibinfo{year}{1983}).

\bibitem[{\citenamefont{de~los Arcos
  et~al.}(2004{\natexlab{b}})\citenamefont{de~los Arcos, Garnier, Oelhafen,
  Mathys, Seo, Domingo, Garcia-Ramos, and Sanchez-Cortes}}]{dlArcosCarbon}
\bibinfo{author}{\bibfnamefont{T.}~\bibnamefont{de~los Arcos}},
  \bibinfo{author}{\bibfnamefont{M.~G.} \bibnamefont{Garnier}},
  \bibinfo{author}{\bibfnamefont{P.}~\bibnamefont{Oelhafen}},
  \bibinfo{author}{\bibfnamefont{D.}~\bibnamefont{Mathys}},
  \bibinfo{author}{\bibfnamefont{J.~W.} \bibnamefont{Seo}},
  \bibinfo{author}{\bibfnamefont{C.}~\bibnamefont{Domingo}},
  \bibinfo{author}{\bibfnamefont{J.~V.} \bibnamefont{Garcia-Ramos}},
  \bibnamefont{and}
  \bibinfo{author}{\bibfnamefont{S.}~\bibnamefont{Sanchez-Cortes}},
  \bibinfo{journal}{Carbon} \textbf{\bibinfo{volume}{42}}, \bibinfo{pages}{187}
  (\bibinfo{year}{2004}{\natexlab{b}}).

\bibitem[{\citenamefont{Moulder et~al.}(1992)\citenamefont{Moulder, Stickle,
  Sobol, and Bomben}}]{XPSHandbook}
\bibinfo{author}{\bibfnamefont{J.~F.} \bibnamefont{Moulder}},
  \bibinfo{author}{\bibfnamefont{W.~F.} \bibnamefont{Stickle}},
  \bibinfo{author}{\bibfnamefont{P.~E.} \bibnamefont{Sobol}}, \bibnamefont{and}
  \bibinfo{author}{\bibfnamefont{K.~D.} \bibnamefont{Bomben}},
  \emph{\bibinfo{title}{Handbook of X-ray Photoelectron Spectroscopy}}
  (\bibinfo{publisher}{Perkin-Elmer}, \bibinfo{address}{Eden Prairie, MN},
  \bibinfo{year}{1992}).

\bibitem[{\citenamefont{Kononchuk et~al.}(1998)\citenamefont{Kononchuk,
  Korablev, Yarykin, and Rozgonyi}}]{KononchukAPL}
\bibinfo{author}{\bibfnamefont{O.}~\bibnamefont{Kononchuk}},
  \bibinfo{author}{\bibfnamefont{K.~G.} \bibnamefont{Korablev}},
  \bibinfo{author}{\bibfnamefont{N.}~\bibnamefont{Yarykin}}, \bibnamefont{and}
  \bibinfo{author}{\bibfnamefont{G.~A.} \bibnamefont{Rozgonyi}},
  \bibinfo{journal}{Applied Physics Letters} \textbf{\bibinfo{volume}{73}},
  \bibinfo{pages}{1206} (\bibinfo{year}{1998}).

\bibitem[{\citenamefont{Liehr et~al.}(1986)\citenamefont{Liehr, Lefakis,
  LeGoues, and Rubloff}}]{LiehrPRB}
\bibinfo{author}{\bibfnamefont{M.}~\bibnamefont{Liehr}},
  \bibinfo{author}{\bibfnamefont{H.}~\bibnamefont{Lefakis}},
  \bibinfo{author}{\bibfnamefont{F.~K.} \bibnamefont{LeGoues}},
  \bibnamefont{and} \bibinfo{author}{\bibfnamefont{G.~W.}
  \bibnamefont{Rubloff}}, \bibinfo{journal}{Physical Review B}
  \textbf{\bibinfo{volume}{33}}, \bibinfo{pages}{5517} (\bibinfo{year}{1986}).

\bibitem[{\citenamefont{Mayer et~al.}(1992)\citenamefont{Mayer, Lin, and
  Garfunkel}}]{MayerSurfSci}
\bibinfo{author}{\bibfnamefont{J.~T.} \bibnamefont{Mayer}},
  \bibinfo{author}{\bibfnamefont{R.~F.} \bibnamefont{Lin}}, \bibnamefont{and}
  \bibinfo{author}{\bibfnamefont{E.}~\bibnamefont{Garfunkel}},
  \bibinfo{journal}{Surface Science} \textbf{\bibinfo{volume}{265}},
  \bibinfo{pages}{102} (\bibinfo{year}{1992}).

\bibitem[{\citenamefont{Conforto and Schmid}(2001)}]{ConfortoPhilMagA}
\bibinfo{author}{\bibfnamefont{E.}~\bibnamefont{Conforto}} \bibnamefont{and}
  \bibinfo{author}{\bibfnamefont{P.~E.} \bibnamefont{Schmid}},
  \bibinfo{journal}{Philosophical Magazine A} \textbf{\bibinfo{volume}{81}},
  \bibinfo{pages}{61} (\bibinfo{year}{2001}).

\bibitem[{\citenamefont{Fedorovich}(1980)}]{FedorovichSovPhysSS}
\bibinfo{author}{\bibfnamefont{N.~A.} \bibnamefont{Fedorovich}},
  \bibinfo{journal}{Soviet Physics Solid State} \textbf{\bibinfo{volume}{22}},
  \bibinfo{pages}{1093} (\bibinfo{year}{1980}).

\bibitem[{\citenamefont{Ohno et~al.}(2003)\citenamefont{Ohno, Iwatsuki,
  Hiraoka, Okazaki, Kishimoto, Maezawa, Shinohara, and Mizutani}}]{OhnoJJAP}
\bibinfo{author}{\bibfnamefont{Y.}~\bibnamefont{Ohno}},
  \bibinfo{author}{\bibfnamefont{S.}~\bibnamefont{Iwatsuki}},
  \bibinfo{author}{\bibfnamefont{T.}~\bibnamefont{Hiraoka}},
  \bibinfo{author}{\bibfnamefont{T.}~\bibnamefont{Okazaki}},
  \bibinfo{author}{\bibfnamefont{S.}~\bibnamefont{Kishimoto}},
  \bibinfo{author}{\bibfnamefont{K.}~\bibnamefont{Maezawa}},
  \bibinfo{author}{\bibfnamefont{H.}~\bibnamefont{Shinohara}},
  \bibnamefont{and} \bibinfo{author}{\bibfnamefont{T.}~\bibnamefont{Mizutani}},
  \bibinfo{journal}{Japanese Journal of Applied Physics}
  \textbf{\bibinfo{volume}{42}}, \bibinfo{pages}{4116} (\bibinfo{year}{2003}).

\bibitem[{\citenamefont{Ghoshtagore}(1969)}]{GhoshJAP}
\bibinfo{author}{\bibfnamefont{R.~N.} \bibnamefont{Ghoshtagore}},
  \bibinfo{journal}{Journal of Applied Physics} \textbf{\bibinfo{volume}{40}},
  \bibinfo{pages}{4374} (\bibinfo{year}{1969}).

\bibitem[{\citenamefont{Liu et~al.}(2004)\citenamefont{Liu, Fang, Lu, Ma,
  Zhang, Yang, Jin, and Gu}}]{LiuJPCB}
\bibinfo{author}{\bibfnamefont{L.~W.} \bibnamefont{Liu}},
  \bibinfo{author}{\bibfnamefont{J.~H.} \bibnamefont{Fang}},
  \bibinfo{author}{\bibfnamefont{L.}~\bibnamefont{Lu}},
  \bibinfo{author}{\bibfnamefont{Y.~J.} \bibnamefont{Ma}},
  \bibinfo{author}{\bibfnamefont{Z.}~\bibnamefont{Zhang}},
  \bibinfo{author}{\bibfnamefont{H.~F.} \bibnamefont{Yang}},
  \bibinfo{author}{\bibfnamefont{A.~Z.} \bibnamefont{Jin}}, \bibnamefont{and}
  \bibinfo{author}{\bibfnamefont{C.~Z.} \bibnamefont{Gu}},
  \bibinfo{journal}{Journal of Physical Chemistry B}
  \textbf{\bibinfo{volume}{108}}, \bibinfo{pages}{18460}
  (\bibinfo{year}{2004}).

\bibitem[{\citenamefont{Backhaus-Ricoult}(1992)}]{BackhausActaMeM}
\bibinfo{author}{\bibfnamefont{M.}~\bibnamefont{Backhaus-Ricoult}},
  \bibinfo{journal}{Acta Metallurgica Et Materialia}
  \textbf{\bibinfo{volume}{40}}, \bibinfo{pages}{S95} (\bibinfo{year}{1992}).

\bibitem[{\citenamefont{Pan and Baptista}(1996)}]{PanJACeramS}
\bibinfo{author}{\bibfnamefont{Y.}~\bibnamefont{Pan}} \bibnamefont{and}
  \bibinfo{author}{\bibfnamefont{J.~L.} \bibnamefont{Baptista}},
  \bibinfo{journal}{Journal of the American Ceramic Society}
  \textbf{\bibinfo{volume}{79}}, \bibinfo{pages}{2017} (\bibinfo{year}{1996}).

\bibitem[{\citenamefont{Hafner et~al.}(2001)\citenamefont{Hafner, Cheung,
  Oosterkamp, and Lieber}}]{HafnerJPCB}
\bibinfo{author}{\bibfnamefont{J.~H.} \bibnamefont{Hafner}},
  \bibinfo{author}{\bibfnamefont{C.~L.} \bibnamefont{Cheung}},
  \bibinfo{author}{\bibfnamefont{T.~H.} \bibnamefont{Oosterkamp}},
  \bibnamefont{and} \bibinfo{author}{\bibfnamefont{C.~M.}
  \bibnamefont{Lieber}}, \bibinfo{journal}{Journal of Physical Chemistry B}
  \textbf{\bibinfo{volume}{105}}, \bibinfo{pages}{743} (\bibinfo{year}{2001}).

\bibitem[{\citenamefont{Shirley}(1972)}]{ShirleyPRB}
\bibinfo{author}{\bibfnamefont{D.~A.} \bibnamefont{Shirley}},
  \bibinfo{journal}{Physical Review B} \textbf{\bibinfo{volume}{5}},
  \bibinfo{pages}{4709} (\bibinfo{year}{1972}).

\end{thebibliography}
\end{document}